\documentstyle[12pt]{article}

\def\b{\beta}

\def\d{\delta}

\def\f{\phi}

\def\l{\lambda}

\def\cu{{\cal U}}

\def\dag{\dagger}
\def\pro{\propto}
\def\la{\left}
\def\ra{\right}
\def\pa{\partial}

\def\inf{\infty}

\def\Hat#1{\rlap{\kern.10em$\widehat{\phantom G}$}#1}
\def\HAt#1{\rlap{\kern.05em$\widehat{\phantom G}$}#1}

\def\cap#1{\rlap{\kern.1em$\widehat{\phantom{G\vrule height.8em}}$}#1{}}
\def\Cap#1{\rlap{\kern.05em$\widehat{\phantom{G\vrule height.8em}}$}#1{}}

\def\VEV#1{\left\langle #1\right\rangle}

\catcode`@=11
\def\underline#1{\relax\ifmmode\@@underline#1\else
        $\@@underline{\hbox{#1}}$\relax\fi}
\catcode`@=12

\def\PRL{Phys. Rev. Lett.\ }

\def\PLB{Phys. Lett. B\ }

\def\NPB{Nucl. Phys. B\ }
\def\IJMPA{Int. J. Mod. Phys. A\ }

\def\ba{\begin{array}}
\def\ea{\end{array}}
\def\be{\begin{equation}}
\def\ee{\end{equation}}
\def\bdm{\begin{displaymath}}
\def\edm{\end{displaymath}}
\def\bea{\begin{eqnarray}}
\def\eea{\end{eqnarray}}
\def\nl{\nonumber \\}

\def\by{\over}
\def\lbl{\label}
\def\sp{~~~~~}
\def\vf{\varphi}

\def\au#1{\\ \vglue .2in $^{#1}$ {\it Department of Physics\\
          University of Arizona, Tucson, AZ 85721}}

\def\cu#1{\\ \vglue .2in $^{#1}$ {\it Department of Physics\\
          University of Colorado, Boulder, CO 80309}}

\def\bt#1#2#3#4#5
        {\begin{titlepage}
         \centerline{\hfill {#1}}
         \centerline{\hfill {#2}}
         \centerline{\hfill {#3}}
         \begin{center}\vskip .2in
         {\large\bf {#4}}\\[.3in]
         {\bf {#5}}\\[.3in]
         {\bf ABSTRACT}\\
         \end{center}
         \begin{quotation}}
\def\et
         {\end{quotation}
          \end{titlepage}}

\newcounter{sxn}

\newcounter{axn}

\def\br{\begin{thebibliography}{abc}}
\def\er{\end{thebibliography}}
\def\rf{\bibitem}

\begin{document}

\bt{COLO-HEP/295}{AZPH-TH/92-43}{December 1992}{Double Scaling Limit
of Scalar Theories on the Lattice}{B. S. Balakrishna \cu{} \\ \vglue
.2in and
\au{}}

Scalar field theories regularized on a $D$ dimensional lattice are
found to exhibit double scaling for a class of critical behaviors
labeled by an integer $m\geq 2$. The continuum theory reached in the
double scaling limit defines a universality class and is of a massless
scalar with only a $m+1$ point self-interaction, but in the presence
of a constant source. The upper critical dimension for this to occur
is the same as that for renormalizability of the $m+1$ point
interaction.

\et

$1/N$ expansion is a powerful tool that has enabled one to study some
of the phenomenon nonperturbative in the coupling constants.  However,
in most cases, it turns out to be possible to study only the leading
term of the $1/N$ series, the large $N$ limit.  One way to examine the
subleading terms is to look for their enhancement that maintains the
virtues of the large $N$ limit. This was possible in the past for
matrix models simulating two dimensional quantum gravity and for some
low dimensional string theories\cite{mat}. There one observed that the
higher order terms of the $1/N$ series, receiving contributions from
higher genus Riemann surfaces, diverged as one approached a critical
point. Remarkably, it was found possible to make use of this
divergence to overcome the $1/N$ suppression and elevate all the terms
of the $1/N$ series to the same level. This approach, called double
scaling, helped one to sum the $1/N$ series to obtain a
nonperturbative solution.

This phenomenon is found to occur in other situations as
well\cite{vec}, as for instance in a class of ${\rm O}(N)$ vector
models.  Here, the presence of double scaling does not in general help
one to sum the $1/N$ series, but provides new insights into the
problem not available otherwise.  Most of these models can be reduced
to a scalar field theory with $1/N$ playing the role of a loop
expansion parameter. The question then arises as to whether double
scaling is more generic and is a phenomenon of scalar theories in
general.  As we will see in this letter, the answer to this question
is yes. A wide class of scalar field theories are found to exhibit
double scaling as the coupling constants approach a critical point.
This occurs for a class of critical behaviors labeled by an integer
$m\geq 2$. In the double scaling limit, one approaches the theory of a
massless scalar field with a $m+1$ point self-interaction in the
presence of a constant source, defining a universality class reached
in the infrared. For even values of $m$, this theory has problems as
the potential is unbounded from below.  For odd values of $m$, one
expects it to be well defined. The short distance theory is not unique
due to universality and the one studied here is a lattice model in $D$
dimensions. The double scaling limit offers an interesting way of
approaching the continuum from the lattice. The upper critical
dimension turns out to be the same as that for renormalizability of a
$m+1$ point interaction

Consider a model given by the following action defined on a $D$
dimensional lattice:
\be
S ~=~ N\sum_{\VEV{xy}}\f(x)(\f(x)-\f(y)) + N\sum_xV(\f(x),\b),
                                                        \lbl{act}
\ee
where $\f(x)$ is a real scalar field. For convenience, we stick to one
scalar, but the conclusions are extendable to cover more fields.  The
lattice sites are labeled by $x$ and the links by $\VEV{xy}$.  $1/N$
plays the role of a loop expansion paraemter, like the Planck's
constant. The potential $V$ may depend on a set of coupling constants
collectively denoted by $\b$. The theory studied here is on the
lattice but the conclusions are equally applicable to those
regularized with a momentum space cutoff. Double scaling is however an
interesting way to approach the continuum from the lattice.

The model is easily solved for large $N$. This is because the
partition function is then dominated at the saddle point governed by a
translationally invariant $\f$. Setting the $\f(x)$'s to be equal to
$\l$ gives a potential $V(\l,\b)$ whose saddle point equation is
\be
V'(\l,\b) ~=~ 0.
\ee
Here a prime denotes differentiation with respect to $\l$. With a
solution giving an absolute minimum, $V(\l,\b)$ gives the free energy
per site for large $N$.

Subleading terms of the $1/N$ series are better examined in momentum
space with the help of Feynman diagrams. Because $1/N$ is a loop
expansion paramter, a graph having $L$ loops is generated at order
$N^{1-L}$.  Starting at an absolute minimum of the potential $V$, the
interaction potential is obtained by translating $\f$ by an amount
$\l$,
\be
V(\f+\l,\b) ~=~ V(\l,\b) + \sum_{p=2}^{\inf}V^p(\l,\b){\f^p\by p!}.
\ee
This shows that a vertex of order $p$ in a Feynman graph has strength
$-V^p(\l,\b)$. $V^2(\l,\b)$ is the mass squared term. The lattice
propagator is $K^{-1}$ where $K$ is the matrix appearing in the
kinetic part of the action (\ref{act}). $K$, with its indices labeling
the sites, has the value $2D$ along the diagonal and $-1$ for all the
$2D$ neighbors. It is diagonalizable in momentum space. Its eigenvalue
along a momentum vector $k$ is $2\sum_i(1-{\rm cos}(k_ia))$, where $i$
runs over all the $D$ components of $k$ and $a$ is the lattice
spacing. For small momenta, this starts off as $k^2a^2$ where a sum
over components is implicit.

Double scaling is a phenomenon near the critical point. One expects in
general a critical behavior to be present for large $N$ itself. In the
space of coupling constants $\b$, there exists in general a class of
critical behaviors. The potential $V(\l,\b)$, with $\l$ treated as a
variable, leads to a definition of $m-$criticality,
\be
V^l(\l_c,\b_c) ~=~ 0, \sp l~=~1,\cdots,m, \sp m\geq 2,
\ee
where a superscript $l$ denotes $l-$th derivative. $\l_c$ and $\b_c$
are the critical values of $\l$ and $\b$ respectively. The potential
$V(\l,\b)$ near $\b_c$ is approximately
\be
V(\l,\b) ~=~ V_c + (\b-\b_c)\pa_\b V_c + (\b-\b_c)\pa_\b V'_c(\l-\l_c) +
{V^{m+1}_c\by(m+1)!}(\l-\l_c)^{m+1},
\ee
where a subscript $c$ denotes evaluation at the critical point.
Because $\l_c$ corresponds to an absolute minimum of $V$, $V^{m+1}_c$
is positive. A summation over all the coupling constants is implicit
in $(\b-\b_c)\pa_\b$. We assume that the critical value $\b_c$ is
approached along $\pa_\b V'_c$ from below.  Using the above
approximation in the saddle pont equation gives
\be
\l-\l_c ~=~ \la[{m!\by V^{m+1}_c}(\b_c-\b)\pa_\b V'_c\ra]^{1/m}.
\ee
This is nonanalytic as $\b\to\b_c$ as is usual in critical phenomenon.
We are interested in $V^p(\l)$ for $p\geq 2$ near the critical point,
\be
V^p(\l,\b) ~=~ {V^{m+1}_c\by(m-p+1)!}\la[{m!\by
V^{m+1}_c}(\b_c-\b)\pa_\b V'_c\ra]^{1-(p-1)/m}, \sp p\leq m+1.
                                                         \lbl{ppv}
\ee
Higher derivatives of $V$ tend to a constant near the critical point.
As we will see later, this results in the suppression of vertices of
order $p>m+1$ in the double scaling limit.

As an example, consider the following potential:
\be
V(\f,\b) = \b e^{2\f} + e^{\f} - \f.
\ee
The saddle point equation is quadratic in $e^{\f}$ leading to the
solution,
\be
e^{\f} = {1\by 4\b}\la(-1+\sqrt{1+8\b}\ra),
\ee
where the other root is ignored as it leads to an unphysical value of
$\f$. Nonanalyticity sets in as $\b\to-1/8$. Because the potential is
unbounded from below for negative $\b$, this is to be viewed as an
analytical continuation. The critical point $\b_c=-1/8$ is found to be
of order $m=2$.

Eq. (\ref{ppv}) shows that the vertices of a Feynman graph of order
$m+1$ and lower go to zero as $\b\to\b_c$. But, we are looking for an
enhancement of the $1/N$ suppressions. This comes from the infrared
divergences that arise because the mass squared term $V^2(\l,\b)$ also
vanishes in this limit. To pick up these contributions, one needs to
take the lattice spacing $a$ to zero as well. The lattice propagator
including the mass squared term is $[K+V^2(\l,\b)]^{-1}$. Because $K$
is of order $a^2$ as $a\to 0$, we require the mass squared term to be
also be of order $a^2$. This determines the rate of approach to the
critical point,
\be
\b_c-\b ~\sim~ a^{2m/(m-1)}.                         \lbl{bto}
\ee
Even if one has begun the discussion with a fixed lattice spacing,
there is a need to scale all the momentum variables, $k\to ka$, in a
Feynman graph by a small parameter $a$ to blow up the infrared region.
Either way, each of the propagators now contributes a factor $1/a^2$
and each of the loops a factor $a^D$. A vertex of order $p$ for $p\leq
m+1$ supplies a factor $a^{2(m-p+1)/(m-1)}$. Let us exclude all
vertices of order $m+2$ and higher for the moment. Also, let us
restrict our attention to graphs having no external legs. If there are
$V_p$ vertices of order $p$ and $E$ propagators in a graph having $L$
loops, the product of $a-$factors is
\be
a^{DL-2E}\Pi_{p=3}^{m+1}\la[a^{2(m-p+1)/(m-1)}\ra]^{V_p} ~=~
a^D\la[a^{D_c-D}\ra]^{1-L},                          \lbl{tas}
\ee
where $D_c$ is the critical dimension,
\be
D_c ~=~ 2\la({m+1\by m-1}\ra).
\ee
Ignoring the overall factor $a^D$, note that this diverges as $a\to 0$
when $D$ is less than the critical dimension $D_c$. It is now clear
how to enhance the $1/N$ suppression factors.  Letting
\be
N ~\sim~ a^{D-D_c},                                       \lbl{nto}
\ee
one can elevate all the loop graphs to the same order with an overall
factor $a^D$. The scaling given by (\ref{bto}) and (\ref{nto}) as
$a\to 0$ is what is meant by double scaling in this letter. The
overall factor $a^D$ is what is needed to relate the free energy per
site to a free energy density. Vertices of order $m+2$ and higher do
not contribute any $a-$factors. In a graph having such vertices, to
obtain the previous result (\ref{tas}), we need to have the factor
$a^{2(m-p+1)/(m-1)}$ for each vertex of order $p\geq m+2$ as well.
Allowing for such factors is equivalent to having
$a^{-2(m-p+1)/(m-1)}$ for each vertex of order $p\geq m+2$ over and
above the result (\ref{tas}). But, $a^{-2(m-p+1)/(m-1)}$ vanishes as
$a\to 0$ at least as fast as $a^{2/(m-1)}$ for $p\geq m+2$. This shows
that all the graphs having vertices of order $m+2$ or higher are
suppressed in the double scaling limit. In the presence of external
legs, the external momenta should be scaled by $a$ just like the
internal ones, which results in an enhancement as before.

The continuum theory one approaches in the double scaling limit is
easily established. It is determined by the following potential for a
continuum field variable $\vf$:
\bea
\sum_{p=2}^{m+1}V^p(\l,\b){\vf^p\by p!} &\sim&
{V^{m+1}_c\by(m+1)!}\la[(\vf+\l-\l_c)^{m+1}-(m+1)(\l-\l_c)^m\vf\ra]
\nl &\to& {V^{m+1}_c\by(m+1)!}\vf^{m+1} - t\vf,          \lbl{uni}
\eea
where $t=(\b_c-\b)\pa_\b V'_c$ and the factors of $a$ are not included
as they are already taken care of. Also, some constants have been
dropped and a translation of $\vf$ has been made in the end. Vertices
of order $p>m+1$ are absent as they are suppressed in the double
scaling limit.  What we have reached is the theory of a massless
scalar field $\vf$ with only a $m+1$ point self-interaction, but in
the presence of a constant source $t$.  Because $V^{m+1}_c$ is
positive, this leads to a well-defined potential for odd values of
$m$. It is problematic for even values of $m$ as it is unbounded from
below. In this case, the double scaling limit is expected to involve
some analytical continuation as in the case of the example mentioned
earlier. This can make the absolute minimum we started with only an
extremum of the potential. The double scaling limit may still exist as
an analytical continuation of the sum of Feynman diagrams.

To justify the analysis based on $1/N$ requires $D$ to be less than
its critical value $D_c=2(m+1)/(m-1)$. $D_c$ takes values $6,4,3$ and
$2$ respectively for $m=2,3,5$ and $\inf$. Because this is also the
critical dimension for requiring renormalizability of a $m+1$ point
interaction, one is in the superrenormalizable regime. The combination
$g=a^{D-D_c}/N$ survives to play the role of a loop expansion
parameter. In other words, $g^{(m-1)/2}$ is the strength of the $m+1$
point interaction.

Because the double scaling limit takes one to the infrared region, the
resulting theory is expected to be highly universal. Already, the
previous arguments have suggested that the theory (\ref{act}) governed
by a general potential $V$ falls to one of the classes given by
(\ref{uni}). But, there is more to this. If the propagator is more
general, say $[Kf(K)]^{-1}$ ignoring the mass squared contribution,
$f(K)$ will tend to a constant $f(0)$ (assumed finite) in the infrared
and the propagator is $\pro K^{-1}$ as before. This is an indication
of the well-known fact that small distance effects like the lattice
structure are irrelevant in the continuum limit. Further, if there are
derivative couplings (involving lattice derivatives), they contribute
at least a factor $k^2\to k^2a^2\sim a^2$.  But this vanishes as $a\to
0$ faster than a $p-$ point vertex strength $V^p(\l,\b)\sim
a^{2(m-p+1)/(m-1)}$ (or constant). Derivative couplings are thus
expected to be suppressed in the double scaling limit. Universality,
being an inherent feature of critical phenomenon in general, is thus
found to be present here as well.

Because of the survival of all vertices of order less than $m+1$ in
the double scaling limit, like for instance the cubic vertex, there
are tadpoles that need to be summed. As shown earlier, a translation
of the field variable removes all the vertices except that of order
$m+1$, but introduces a constant source $t$. The resulting theory is
easier to handle in the effective action approach. Because the source
is a constant, only the effective potential matters. The source $t$ is
now traded for the expectation value $v$ of the continuum field
variable $\vf$. In this approach only the one particle irreducible
diagrams are retained, thus excluding the tadpoles. The effective
potential computed this way gives us the one we are looking for after
the addition of the source term $-tv$. Minimization of the resulting
effective potential relates $v$ and $t$. Because the loop expansion
parameter $g$ is dimensionful having mass dimension $D_c-D$, and $v$
has mass dimension $2/(m-1)$ in our conventions, an expansion in loops
is actually a perturbation in $g/v^{(D_c-D)(m-1)/2}$. The tree level
potential suggests that $v\pro t^{1/m}$, hence this is a perturbation
in $g/t^{(D_c-D)(m-1)/(2m)}$.  Ultraviolet divergences that might
appear in the double scaling limit can be handled in the usual way by
the addition of counter terms to the bare lattice action we started
with.

One way to generalize the conclusions for more than one scalars is as
follows. The variables $\f$ and $\l$ are now given an index to label
the various scalars. Products of $\f$ and $\l$, and various
derivatives of $V$, receive more indices like tensors. $m-$criticality
could be defined by requiring all the derivatives of $V$ of order less
than $m+1$ to vanish. Let us assume that a solution to the saddle
point equation giving an absolute minimum exists near the critical
point. Now, the order of magnitudes of mass and the vertex strengths
are all the same as before. All the scalars help in double scaling and
hence survive. The final result is the theory of massless scalars with
only $m+1$ point interactions, but in the presence of a constant
source. The source points along a direction given by $\b\pa_\b V'_c$.
A simple example is given by an O$(M)$ theory where the field variable
$\f$ has $M$ components. The potential $V$ is now a function of
$\f^\dag\f$, the square of the length of the $M-$vector $\f$. It is
easily found that the double scaling limit of this theory is the same
as before, that of one massless scalar with $m+1$ point
self-interaction in the presence of a constant source, except for the
presence of $M-1$ additional free massless scalars.

So far, we had treated the coupling parameters as constants. It is
possible to generalize the results to cover space dependent couplings.
The conclusions are the same, except that the source $t$ is now
replaced by $t(x)$. Space dependent source helps in deriving the
Schwinger-Dyson equation. For the $m+1$ theory, this is easily
obtained,
\be
{V_c^{m+1}\by m!}g^m\d_{t(x)}^mZ - g\pa^2\d_{t(x)}Z - t(x)Z = 0,
\ee
where $Z$ is the functional integral of the continuum theory and
$t(x)=(\b_c-\b(x))\pa_\b V'_c$. $\d_{t(x)}$ is the functional
derivative with respect to $t(x)$. This differential equation is what
one expects to emerge in the double scaling limit from an analogous
one for the original theory for any $m\geq 2$. In zero dimension, the
term involving the spatial derivative is absent and the functional
derivatives are ordinary derivatives. Then, the solution to the
Schwinger-Dyson equation for $m=2$ is given in terms of modified
Bessel functions,
\be
Z ~\pro~ t^{1/2} I_{1/3}\la(2t^{3/2}/3\ra),
\ee
where $t$ has been scaled to $(V^3_cg^2/2)^{1/3}t$. This sums the
series generated by the $1/N$ expansion in the double scaling limit.
The asymptotic expansion of the Bessel function for large values of
$t$ recovers the series. Though the potential is unbounded from below
for $m=2$, the Schwinger-Dyson equation admits a solution that agrees
with the Feynman diagrams in the double scaling. This is because one
expects the double scaling limit to involve some analytical
continuation for even $m$

In this letter, it is argued that the phenomenon of double scaling
exists in a wide class of scalar field theories. It is discussed here
for theories regularized on a $D$ dimensional lattice. The critical
behaviors involved here belong to a class ordered by an integer $m\geq
2$. The continuum theory reached in the double scaling limit turns out
to be the theory of a massless scalar with only a $m+1$ point
self-interaction, but in the presence of a constant source. This
provides an interesting way to approach the continuum limit from the
lattice. Only the universality class given by the $m+1$ theory is
expected to matter in the double scaling limit. The arguments are
expected to hold for dimensions less than the critical dimension for
renormalizability of a $m+1$ point interaction.  Besides being
interesting in its own right, this work might turn out to be useful in
the analysis of double scaling for the scalars that are expected to
emerge in the $1/N$ expansion of various theories, like for instance
the O$(N)$ models\cite{vec} or those constructed to induce
QCD\cite{qcd}.

I would like to thank Professor Anna Hasenfratz for discussions. This
work is supported by NSF Grant PHY-9023257.

\br
\rf{mat} E. Br\'{e}zin and V. A. Kazakov, \PLB 236, 144 (1990);\\
M. R. Douglas and S. H. Shenker, \NPB 335, 635 (1990);\\ D. J. Gross
and A. A. Migdal, \PRL 64, 127 (1990); \NPB 340, 333 (1990).
\rf{vec} J. Ambjorn, B. Durhuus and T. J\'{o}nsson, \PLB 244, 403 (1990);\\
S. Nishigaki and T. Yoneya, \NPB 348, 787 (1991);\\ A. Anderson, R. C.
Myers and V. Periwal, \PLB 254, 89 (1991); \NPB 360, 463 (1991);\\ P.
Di Vecchia, M. Kato and N. Ohta, \NPB 357, 495 (1991); \IJMPA 7, 1391
(1992);\\ J.  Zinn-Justin, \PLB 257, 335 (1991).
\rf{qcd} V. A. Kazakov and A. A. Migdal, preprint PUPT-1322,
LPTENS-92/15, to be published in \NPB;\\ A. A. Migdal, preprints
PUPT-1323; PUPT-1332; LPTENS-92/22;\\ I. I. Kogan, G.  W. Semenoff and
N. Weiss, preprint UBCTP-92-022;\\ D. Gross, PUPT-1335.
\er

\end{document}